%% file: main.tex
\documentclass[sigconf,natbib=false,balance=false,nonacm]{acmart}
\AtBeginDocument{%
  }

\setcopyright{acmlicensed}
\copyrightyear{2027}
\acmYear{2027}
\acmDOI{XXXXXXX.XXXXXXX}
\acmConference[HSCC+ICCPS '27]
{ACM/IEEE International Conference on Cyber-Physical and Autonomous Systems}
{May 17--20, 2027}
{Boulder, CO, USA}
\acmISBN{978-1-4503-XXXX-X/2018/06}

\RequirePackage[
  datamodel=acmdatamodel,
  style=acmnumeric,
    sorting=none,
  sortcites=true,
  ]{biblatex}

\usepackage{acronym}
\usepackage{tikz}
\usetikzlibrary{positioning, patterns, patterns.meta, calc}
\usetikzlibrary{patterns, patterns.meta}
\usepackage{todonotes}
\usepackage{subcaption}

\usepackage{custom_colors}

\begin{document}

\include{definitions}

\title{Keeping Shielded Data-Driven Controllers Safe Under System Evolution via Model Consistency}

\author{Roland Stolz}
\email{roland.stolz@tum.de}
\affiliation{%
  \institution{Technical University of Munich}
  \city{Munich}
  \country{Germany}
}

\author{Vitus Lüntzel}
\email{vitus.luentzel@kit.edu}
\affiliation{%
  \institution{Karlsruhe Institute of Technology}
  \city{Karlsruhe}
  \country{Germany}
}

\author{Romain Pascual}
\email{romain.pascual@centralesupelec.fr}
\affiliation{%
  \institution{Université Paris-Saclay}
  \city{Paris}
  \country{France}
}

\author{Eric Sax}
\email{eric.sax@kit.edu}
\affiliation{%
  \institution{Karlsruhe Institute of Technology}
  \city{Karlsruhe}
  \country{Germany}
}

\author{Matthias Althoff}
\email{althoff@tum.de}
\affiliation{%
  \institution{Technical University of Munich}
  \city{Munich}
  \country{Germany}
}

\begin{abstract}
Cyber-physical systems (CPS) increasingly contain data-driven controllers, which raise safety concerns given their black-box nature. A common mitigation is to enforce system safety via a shield around the controller, which restricts the inputs based on assumptions about the physical system. However, these assumptions are usually fixed and may become invalid as the system evolves (during development or due to component degradation), possibly leading to unsafe behavior. We address this limitation in a model-driven approach that treats both the data-driven controller and safety shield as first-class models that are integrated into a virtual single underlying model (V-SUM) alongside system models.
We describe system evolution as changes in the associated models, such that consistency checking yields a continuous assessment of safety guarantees. In case of inconsistencies, consistency-preservation rules automatically update the models and restore a valid safety shield. We evaluate our approach on a reach-avoid task for a unicycle system.
The results show that consistency-driven evolution effectively maintains safety guarantees under system changes and provides a principled basis for evaluating and restoring performance.
	% The results show that consistency-driven evolution can maintain both safety and performance when physical parameters vary, such as mass and actuation limits.
\end{abstract}

\begin{CCSXML}
<ccs2012>
   <concept>
       <concept_id>10010520.10010553</concept_id>
       <concept_desc>Computer systems organization~Embedded and cyber-physical systems</concept_desc>
       <concept_significance>500</concept_significance>
       </concept>
   <concept>
       <concept_id>10010147.10010257.10010258.10010261</concept_id>
       <concept_desc>Computing methodologies~Reinforcement learning</concept_desc>
       <concept_significance>500</concept_significance>
       </concept>
   <concept>
       <concept_id>10010147.10010178.10010213</concept_id>
       <concept_desc>Computing methodologies~Control methods</concept_desc>
       <concept_significance>300</concept_significance>
       </concept>
   <concept>
       <concept_id>10010147.10010341.10010342.10010343</concept_id>
       <concept_desc>Computing methodologies~Modeling methodologies</concept_desc>
       <concept_significance>100</concept_significance>
       </concept>
 </ccs2012>
\end{CCSXML}

\ccsdesc[500]{Computer systems organization~Embedded and cyber-physical systems}
\ccsdesc[500]{Computing methodologies~Reinforcement learning}
\ccsdesc[300]{Computing methodologies~Control methods}
\ccsdesc[100]{Computing methodologies~Modeling methodologies}

\keywords{Safe reinforcement learning, safety shields, cyber-physical systems, system evolution, consistency management}

\maketitle

\section{Introduction}
\input{sections/introduction.tex}

\section{Preliminaries}\label{sec:preliminary}
\input{sections/preliminaries}

\section{Method}\label{sec:method}
\input{sections/method}

\section{Experiments}\label{sec:experiments}
\input{sections/experiments}

\section{Conclusion}\label{sec:conclusion}

\input{sections/conclusion}

\section*{Acknowledgment}
We gratefully acknowledge that this project was funded by the Deutsche Forschungsgemeinschaft (DFG, German Research Foundation) – SFB 1608 – 501798263.

\printbibliography

\clearpage
\appendix
\section*{Appendix}
\input{sections/appendix}

\end{document}

%% file: definitions.tex
% -- Custom commands --
\newcommand{\mc}[1]{\mathcal{#1}}
\newcommand{\poldot}{\pol{\cdot}}
\newcommand{\polsdot}{\pols{\cdot}}
\newcommand{\safeu}{\mathcal{U}^s}
\newcommand{\safeus}{\mc{U}^s(x)}
\newcommand{\safepol}{\tilde{\pi}_\theta}

\newcommand{\polshield}{\tilde{\pi}_\theta}
\newcommand{\cor}{\mathit{CR}}
\newcommand{\cpr}{\mathit{CPR}}
\newcommand{\Mshield}{M_\text{shield}}
\newcommand{\crshield}{\cor_\text{safe}}
\newcommand{\cprshield}{\cpr_\text{safe}}
\newcommand{\Magent}{M_\text{agent}}
\newcommand{\cragent}{\cor_\text{behav}}
\newcommand{\cpragent}{\cpr_\text{behav}}
\newcommand{\thr}{\tau}
\newcommand{\dt}{\Delta t}

\newcommand*{\customcomment}[3]{%
    \todo[color=#1!30, inline, caption={#2: #3}]{\textbf{#2:} #3}%
}
\newcommand*{\commentVitus}[1]{\customcomment{orange}{Vitus}{#1}}
\newcommand*{\commentRoland}[1]{\customcomment{blue}{Roland}{#1}}
\newcommand*{\commentRomain}[1]{\customcomment{green}{Romain}{#1}}

% -- Acronyms--
\newacro{RL}{reinforcement learning}
\newacro{MDP}{Markov decision process}
\newacro{CPS}{cyber-physical system}
\newacroplural{CPS}[CPS]{cyber-physical systems}
\newacro{V-SUM}{virtual single underlying model}
\newacro{V-SUMM}{virtual single underlying metamodel}
\newacro{PPO}{proximal policy optimization}
\newacro{ODD}{Operational Design Domain}
\newacro{OMG}{Object Management Group}
\newacro{SysML}{Systems Modeling Language}
\newacro{E/E}{Electric/Electronic}
\newacro{MBSE}{Model-based Systems Engineering}
\newacro{CPR}{consistency-preservation rule}
\newacroplural{CPRs}{consistency-preservation rules}

%% file: sections/introduction.tex
With the rising capabilities of machine learning, data-driven controllers, such as deep \ac{RL} agents, are
increasingly considered for complex control tasks in \acp{CPS} \cite{kiran_deep_2022}.
However, their black-box nature does not provide any guarantees on their behavior, which makes them unsuitable for safety-critical applications \cite{alshiekh_2018_safe}.
Safety shields have been proposed to address this limitation \cite{kasaura_benchmarking_2023,krasowski_provably_2023,feng_adaptive_2025}, by overriding unsafe actions of the controller at runtime based on a formal safety specification.

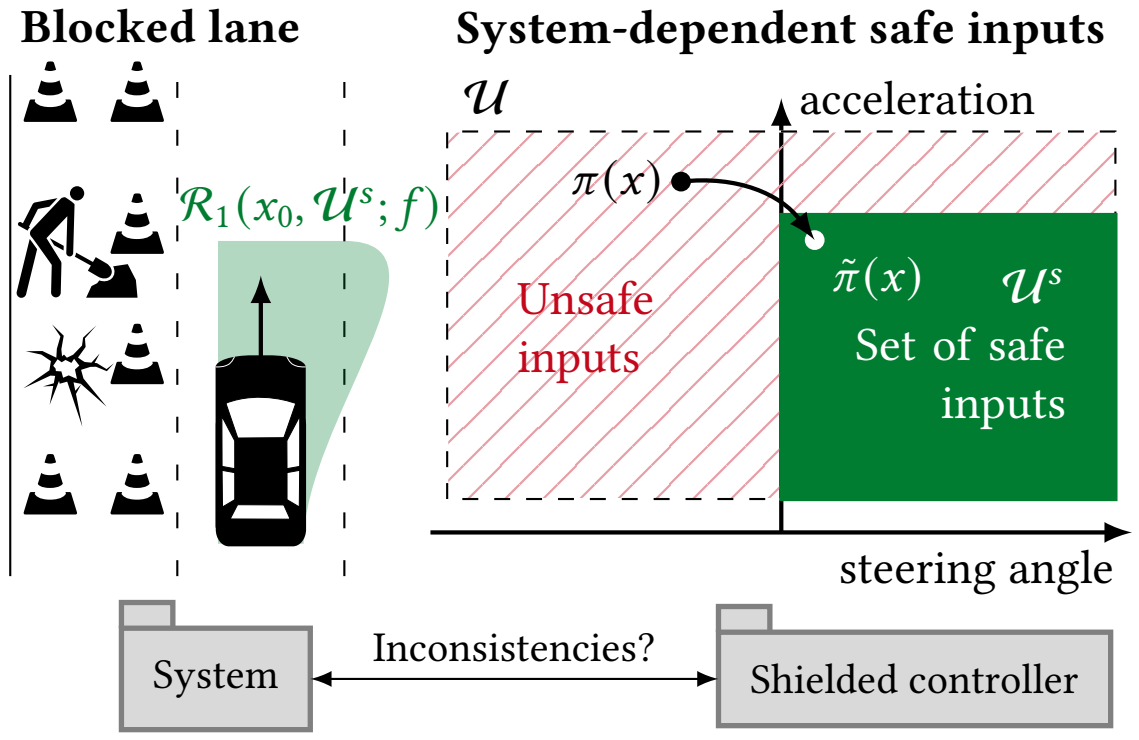
\begin{figure}[tb]
	\centering
	\input{figures/illustrative_example.tex}
	\caption{A data-driven controller proposes an unsafe input \(\pi(x)\), which the safety shield transforms into the safe input \(\tilde{\pi}(x)\). The shaded green area is the region the vehicle can occupy in the future, given the set of safe inputs \(\safeu\), which is computed on assumptions about the system.}
	\label{fig:illustrative_example}
\end{figure}

Figure~\ref{fig:illustrative_example} illustrates an example of the setting considered in this paper. An autonomous vehicle approaches a construction site. The learned controller proposes a steering command that would lead to a collision. Before execution, a safety shield corrects the proposed action. The main challenge is that the shield itself relies on assumptions about the physical system, such as braking capability or tire friction. These assumptions may change during system evolution, e.g., redesign or natural wear, causing the shield to become invalid. 
Similarly, the learned controller was trained under the original dynamics, and its performance may degrade when they shift~\cite{filos_can_2020}.
% Both components are therefore affected by changes in the underlying system models.

Both components can be hardened against such changes individually. For the shield, robust safety filters account for bounded disturbances and model uncertainty~\cite{jankovic_robust_2018}. However, the bounds must be tight to avoid overly conservative behavior, hence they cannot account for large changes to the system.
On the controller side, \ac{RL} methods for non-stationary environments either train policies to be robust against a range of dynamics \cite{moos_robust_2022} or adapt them to the change \cite{padakandla_survey_2021}. In both cases, the components remain isolated from the system models, so a known change to the system is not traced to the shield and controller that depend on it.

Tracing such changes across models is the goal of multi-model consistency management.
Consistency is a weak form of correctness of the \ac{CPS}, i.e., the overlapping system models are free of contradictions and are thus jointly realizable~\cite{spanoudakisInconsistencyManagementSoftware2001}.
Model-driven engineering approaches preserve consistency by relating models through explicit correspondences and propagating changes between them automatically~\cite{klare_commonalities_2019}.
We build upon Vitruvius~\cite{Vitruvius}, which realizes this through the notion of consistency relations and preservation rules across heterogeneous models. 
These approaches require structurally and semantically well-defined models~\cite{knapp_multiview_2018}. A safety shield can be defined this way, but the semantics of a learned controller are only observable through its behavior.
Behavioral notions of consistency based on observable properties relax this requirement~\cite{farber_observable_2026}, but do not explicitly address black-box models.

To ensure consistency of data-driven controllers, we treat both the safety shield and the learned controller as first-class models within a multi-model consistency framework.
For the safety shield, consistency captures the validity of the underlying assumptions on the system and for learned controllers, consistency must be defined at the behavioral level. 
Thus, we distinguish mandatory \emph{safety consistency} for the safety shield and \emph{behavioral consistency} for the data-driven controller. 
% These notions are to be integrated with already available consistency specifications to the other models.
%
% We therefore argue that consistency for such models must be defined at the behavioral level, and introduce the notion of behavioral consistency to enable consistency specifications for models of data-driven controllers within a \ac{V-SUM}. 
% \textcolor{purple}{It is worth noting that the consistency specifications for the safety shield are more important than the behavioral consistency specifications.}
% In this work, we reason over mathematical abstractions of the models, only retaining the relevant parameters and behavior needed for reasoning, while concrete engineering artifacts (e.g., implementations of controllers or simulators) will be later discussed in Sec.~\ref{sec:experiments}.
% We distinguish between two types of consistency that serve different purposes. \emph{Safety consistency} is mandatory to ensure correctness of the system and must always be satisfied. \emph{Performance consistency} captures performance degradation and is therefore treated as a soft constraint. This distinction enables a principled integration of data-driven components into model-driven engineering despite their black-box nature.
% 
We make the following contributions:
\begin{enumerate}
    \item We extend multi-model consistency to learning-enabled \ac{CPS} by treating the safety shield and the data-driven controller as individual first-class models;
    \item We define safety consistency relations for safety shields, and behavioral consistency relations for learned controllers;
    \item We propose corresponding consistency preservation rules to restore consistency under system evolution;
    \item We demonstrate the approach on a reach-avoid task with unicycle dynamics controlled by an \ac{RL} agent.
\end{enumerate}

%% file: figures/illustrative_example.tex
\begin{tikzpicture}[scale=0.85, >= latex]
	\def\umltabw{0.4cm}   % tab width
	\def\umltabh{0.2cm}  % tab height
	\tikzset{
		umlpkg/.style={
				rectangle, draw=#1, line width=1.0,
				fill=#1, fill opacity=0.3, draw opacity=1, text opacity=1,
				align=center, inner sep=6pt, outer sep=0pt, minimum height=0.6cm,
				append after command={
						\pgfextra{%
							\coordinate (umlbl) at (\tikzlastnode.north west);
							\fill[#1, fill opacity=0.3] (umlbl) rectangle ++(\umltabw,\umltabh);
							\draw[line width=1.0, draw=#1, line cap=butt]
							(umlbl) -- ++(0,\umltabh) -- ++(\umltabw,0) -- ++(0,-\umltabh);
						}%
					}
			}
	}

    \node[umlpkg=gray] (system) at (-1.4,-0.8) {\small System};
    \node[umlpkg=gray, right=2.7cm of system.east] (shield) {\small Shielded controller};
    \draw[<->] (system) -- node[above] {\small Inconsistencies?} (shield);
    
	\begin{scope}[shift={(-3, 0)}, scale=1.3]
		\node at (0.9,3.3) {\textbf{Blocked lane}};
        
		% Main rectangular shape
		\draw (0,0) to (0,3);
		\draw[loosely dashed] (1,0) to (1,3);
		\draw[loosely dashed] (2,0) to (2,3);
		%\draw (0.1,0.5) rectangle (0.9,3);

		\begin{scope}[shift={(0.0, -0.1)}]
			\node at (0.24,0.65) {\includegraphics[width=0.4cm]{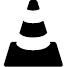}};
			\node at (0.76,0.65) {\includegraphics[width=0.4cm]{figures/assets/cone.pdf}};
			\node at (0.24,3) {\includegraphics[width=0.4cm]{figures/assets/cone.pdf}};
			\node at (0.76,3) {\includegraphics[width=0.4cm]{figures/assets/cone.pdf}};
			\node at (0.76,1.43) {\includegraphics[width=0.4cm]{figures/assets/cone.pdf}};
			\node at (0.76,2.21) {\includegraphics[width=0.4cm]{figures/assets/cone.pdf}};

		\end{scope}
		% \draw[fill, opacity=0.5] (0.1,0.5) rectangle (0.9,3);
		\node at (0.35,1.2) {\includegraphics[width=0.6cm]{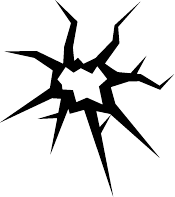}};
		\node at (0.4,2.0) {\includegraphics[width=0.8cm]{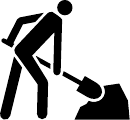}};
		% \draw (1.3,0.3) rectangle (1.7,1.2);

		\filldraw[opacity=0.3, darkgreen] (1.25, 0.2) -- (1.25, 2.0) -- (2.0, 2.0) to[in=90, out=0] (1.75, 0.2) -- cycle;
		\node[darkgreen] at (1.8, 2.2) {\(\mc{R}_1 (x_0, \safeu; f)\)};

		\node at (1.5,0.75) {\includegraphics[width=0.6cm]{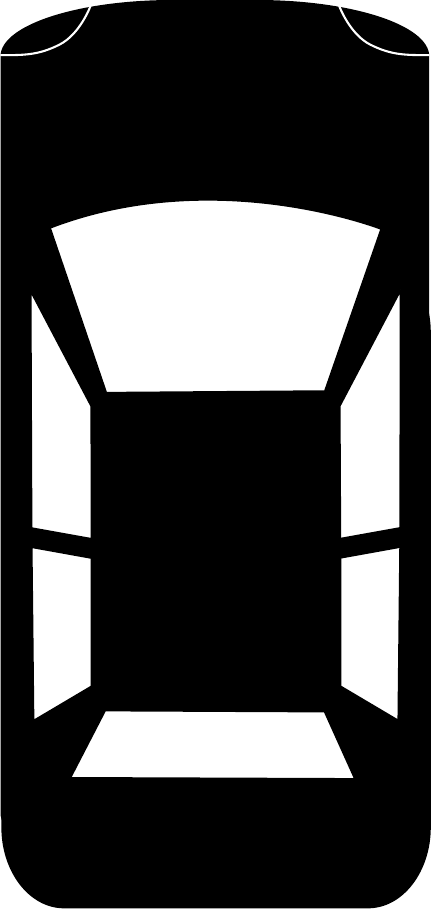}};
		\draw[->,thick] (1.5,1.3) -- (1.5,1.8);
		% \node at (0.9,-0.5) {\textbf{Blocked lane segment}};

	\end{scope}
	% Right side - plot
	\begin{scope}[shift={(3, 0.6)}, scale=1.3]
		\node at (0.0,2.80) {\textbf{System-dependent safe inputs}};
		\draw[->,thick] (-2.1,-0.2) -- (2.1,-0.2) node[below left] {steering angle};
		\draw[->,thick] (0,-0.2) -- (0,2.4) node[right] {acceleration};
		\draw[pattern={Lines[angle=45,distance=5pt]}, pattern color=red!40, draw=white] (-2,0) rectangle (2,2.2);
		\draw[dashed] (-2,0) rectangle (2,2.2);
		\node[above right] at (-2,2.2) {$\mathcal{U}$};
		%\draw[thick] (0,0) rectangle (2,2.2);
		% \draw[thick, fill, opacity=1.0, blue](0,0) -- (0,1.7) -- (1,1.7) -- (2,0) -- cycle;
		\draw[thick, fill, opacity=1.0, darkgreen](0,0) -- (0,1.7) -- (2,1.7) -- (2,0) -- cycle;
		\node[white, text width=2.0cm, align=right] at (0.8,0.9) {$\safeu$\\ Set of safe inputs};
		\node[red, text width=1.5cm] at (-0.9,1.0) {Unsafe inputs};
		% Red dot for pi(x), above right of the unsafe region label
		\filldraw[black] (-0.6,1.9) circle (1.5pt) node[left, black] {$\pi(x)$};
		% White dot near the boundary of safe set, labelled \tilde\pi(x)
		\filldraw[white] (0.2,1.55) circle (1.5pt) node[below right, white] {$\tilde\pi(x)$};
		% Arrow from red dot to white dot
		\draw[->, thick] (-0.6,1.9) to[bend left=30] (0.2,1.55);
		% \node at (0.0,-0.815) {\textbf{System-dependent safe inputs}};
	\end{scope}
\end{tikzpicture}

%% file: sections/preliminaries.tex
We consider a \ac{CPS} represented by a dynamic system model. It is controlled by a data-driven controller that must adhere to given safety specifications. We illustrate these definitions using the running example in Fig.~\ref{fig:illustrative_example}, where the control inputs of a learned controller are constrained with a safety shield under changing system parameters.

\subsection{System Dynamics and Data-Driven Control}\label{sec:dynamics}
The dynamics are represented by a discrete-time system:
\begin{equation}\label{eq:system}
	x_{k+1} = f(x_k, u_k, w_k),
\end{equation}
where $x_k \in \mc{X} \subseteq \mathbb{R}^{n}$ is the state at time step $k$, $u_k \in \mc{U} \subseteq \mathbb{R}^{m}$ is the control input, and $w_k \in \mc{W}$ represents the bounded system noise.
The function \(f\) is derived from the description of the system (its models) and may evolve over time.
Given a state \(x_0\) and a set of inputs \(\tilde{\mc{U}}\), the set of reachable states of \(f\) after one time step is:
\begin{equation}\label{eq:reachable_set}
	\mc{R}_1(x_0, \tilde{\mc{U}}; f) =
	\{ f(x_0, u, w) \mid u \in \tilde{\mc{U}}, w \in \mc{W} \}.
\end{equation}
We consider a parametric controller for the system $\pi_\theta : \mc X \rightarrow \mc{U}$, where $\theta \in \Theta$ are parameters learned from data, e.g., the weights of a neural network.
The policy $\pi_\theta$ is treated as a black-box and only its input-output behavior is known and accessible.
The parameters $\theta$ are optimized with respect to a performance objective $J(\pi_\theta)$.
For instance, in reinforcement learning,
$J$ typically represents the expected cumulative discounted reward, and $\theta$ is updated via gradient ascent through interaction with the environment \cite{sutton_reinforcement_2018}.

\subsection{Shielding of Data-Driven Controllers}\label{sec:pre:acrl}
Since the learned control policies are black boxes, we do not know whether their chosen inputs are safe, and hence safeguard them. We assume the existence of a state-dependent set of safe inputs $\safeus \subseteq \mc{U}$, defined such that applying any $u \in \safeus$ in the state $x$ preserves safety.
Incorporating \(\safeus\) into the controller gives the shielded policy $\safepol: \mc{X} \rightarrow \mc{U}$ that maximizes the objective $J(\safepol)$ while ensuring the selected input belongs to $\safeus$ \cite{krasowski_provably_2023, kasaura_benchmarking_2023, stolz_improving_2026}.
Different methods achieve the containment in $\safeus$, e.g., ray masking \cite{stolz_excluding_2024}, alpha projection \cite{kasaura_benchmarking_2023}, or truncated distributions \cite{stolz_improving_2026}. For simplicity, we use boundary projection, which minimally adapts the input proposed by the data-driven controller \(\pi_\theta(x)\) \cite[Sec.~2.2]{krasowski_provably_2023}:
\begin{equation}\label{eq:boundary_projection}
	\begin{aligned}
		\tilde{\pi}_\theta(x) = & \arg \min_u \quad \Vert \pi_\theta(x) - u \Vert \\
		                        & \text{subject to} \quad u \in \safeus.
	\end{aligned}
\end{equation}

\subsection{Consistency Preservation}
Consistency preservation rules define how a change in one model induces changes in the other models to preserve consistency between all of them, thereby enabling automatic reaction to changes.
For our purposes, we will use the formalization of the \ac{V-SUM} given in~\cite{pascual_formal_2025} that provides a black-box perspective to reason about \mbox{(meta)models} and the associated consistency relation. In those settings, the metamodels are reduced to the sets of their valid instances, and the collection of consistency specifications is aggregated into a global consistency relation \(\cor\). Formally, we consider \(\cor \subseteq M\), where \(M = M_1 \times \ldots \times M_n\) and the \(M_i\)'s are the relevant metamodels. The \ac{V-SUMM} can then be defined as \(\mc{M} = (M, \cor)\).
Further, delta-based consistency preservation rules define a change $\delta_M: M \rightarrow M$ as the modification between two states of a model. This notion is used to define a \ac{CPR} as a function $\cpr: M \times \Delta_M \rightarrow \Delta_M$, where $\Delta_M = \Delta_{M_1} \times \ldots \times \Delta_{M_n}$ and $\Delta_{M_i}$ is the set of all changes for metamodel $M_i$ \cite[Sec.~6.2]{pascual_formal_2025}. Intuitively, given a set of changes of some models, the $\cpr$ defines the necessary changes in the other models to preserve consistency.

%% file: sections/method.tex
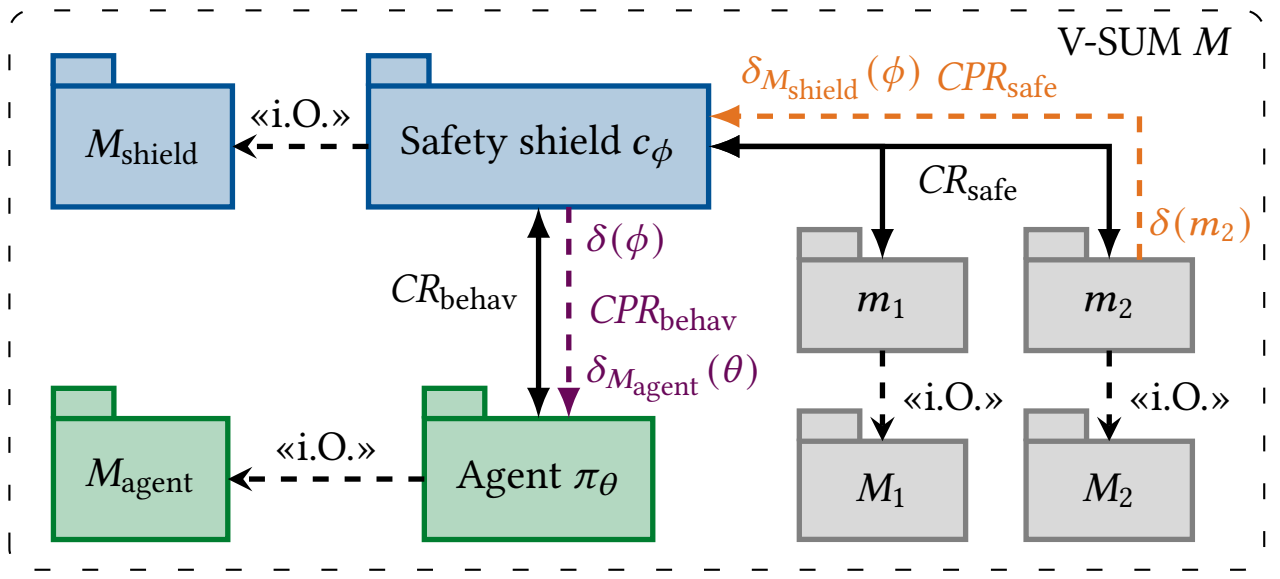
\begin{figure}[tb]
	\centering
	\input{figures/overview.tex}
	\caption{Illustration of our proposed framework for consistency-based change evaluation of shielded data-driven controllers for \ac{CPS} using the \ac{V-SUM} formalization of \cite{pascual_formal_2025}.}
	\label{fig:overview}
\end{figure}

We integrate the shielded data-driven controller into a \ac{V-SUMM} with two metamodels: the safety shield $\Mshield$ in Sec.~\ref{sec:shield_mm}, and the data-driven controller $\Magent$ in Sec.~\ref{sec:agent_mm}.
We assume that the \ac{V-SUMM} $\mc{M} = (M, \cor)$ also contains $k$ other metamodels, hence
\begin{equation}
	M = M_1 \times \cdots \times M_k \times \Mshield \times \Magent.
\end{equation}
Sec.~\ref{sec:consistency_relations} defines the consistency relations $ \cor \subseteq M$, which give us the set of consistent \ac{V-SUM}s, and Sec.~\ref{sec:cprs} defines the consistency preservation rules. We illustrate the proposed framework in Fig.~\ref{fig:overview}.

\subsection{Safety Shield Metamodel}\label{sec:shield_mm}

We construct the safety shield using a set of safe physical states of the system defined by inequality constraints \(h_i(x) \geq 0\). Each constraint encodes a safety-relevant condition on the system state. In our running example, this would be, e.g., keeping the distance to the construction site larger than the safe braking distance, and remaining below a speed limit. Given a collection of $N$ such constraints, the safe set is
\begin{equation}\label{eq:rci_set}
	\mc{X}^s := \{x \in \mc{X} \mid h_i(x) \geq 0,\; \forall i=1,\dots,N\}.
\end{equation}
We can now map the safe state set \(\mc{X}^s\) to the input space via one-step reachability, as defined in \eqref{eq:reachable_set}, thereby obtaining the safety shield models. See Fig.~\ref{fig:illustrative_example} for an illustration.
\begin{definition}[$\Mshield$]\label{def:m_shield}
    Let \(\Phi = \mc{P}(\mathbb{R}^m) \times \mc{F} \times \mc{P}(\mc{X})\) denote the parameter space, where \(\mc{F}\) is the set of admissible dynamics functions. Each parameter \(\phi = (\mc{U},f, \mc{X}^s) \in \Phi\) thus specifies input space, system dynamics and the safe state set. For a given \(\phi \in \Phi\), the corresponding safety shield is \(c_\phi: \mc X \to \mc P(\mc{U})\), defined as
	\begin{equation}
		c_\phi(x):= \safeus = \{ u \in \mc{U} \mid \mc{R}_1(x, \{u\}; f) \subseteq \mc{X}^s \}.
	\end{equation}
    The metamodel \(\Mshield\) is then the set of all such safety shields.
\end{definition}
To guarantee that $c_\phi(x)$ is non-empty for all safe states, $\mc{X}^s$ must be robust control invariant, i.e., for every state $x \in \mc{X}^s$, there exists a control input $u$ that keeps the system within $\mc{X}^s$ for all bounded disturbances. This property holds when $h_i$ are robust control barrier functions \cite{jankovic_robust_2018}.

\subsection{Data-Driven Controller Metamodel}\label{sec:agent_mm}
The data-driven controller is defined by the learned policy, which selects control inputs given the current state of the system.
\begin{definition}[$\Magent$]
    The metamodel $\Magent$ is the set of all policies $\pi_\theta : \mc{X} \rightarrow \mc{U}$, where $\theta \in \Theta$ are parameters encoding the learning agent.
	% The metamodel $\Magent$ is the set of all data-driven control policies $\pi_\theta: \mc{X} \times \mc{Z} \rightarrow \mc{U}$, where the parameters $\theta \in \mathbb{R}^d$ are learned from data.
\end{definition}
We use the policies introduced in Sec.~\ref{sec:dynamics}. The parameter space $\Theta$ abstracts over both the architecture of the neural network and parameters learned from the data, such that policies are treated as black-box input-output mappings.

\subsection{Consistency Relations}\label{sec:consistency_relations}
The safety shield relies on physical properties of the system, such as the input space $\mc{U}$ and the system dynamics, which are (implicitly) specified by other models of the \ac{V-SUM}.
Ensuring consistency therefore requires relating heterogeneous models at a semantic level and cannot be obtained by mere syntactic agreements.

We use observables~\cite{farber_observable_2026} to formalize such semantic consistency.
An observable $o$ defines a function $o_i: M_i \rightarrow \mc{P}(V_o)$ for each metamodel \(M_i\) of the \ac{V-SUMM} that maps a model to a set of admissible values in a shared domain $V_o$. Consistency is then expressed as the existence of a joint realization across models.

\begin{definition}[$\crshield$]\label{def:consistency_shield}
	The consistency relation $\crshield$ holds if
	\begin{equation}\label{eq:consistency_shield}
		% \crshield(m_1, \ldots, m_k, c_\phi) \Longleftrightarrow \phi \subseteq \bigcap_{i=1}^k o_i(m_i),
		\crshield(m_1, \ldots, m_k, c_\phi) \Longleftrightarrow \phi \in \bigcap_{i=1}^k o_i(m_i),
	\end{equation}
    where $o_i : M_i \rightarrow \mc{P}(\Phi)$ maps model instances to admissible parameter sets of the safety shield.
\end{definition}

This formulation ensures that the parameters $\phi$ of the safety shield are compatible with all constraints imposed by the other models, without requiring the models to explicitly define them. 
While observables provide a principled way to define consistency, they rely on system-level properties and the availability of meaningful abstractions thereof. This assumption fails for the data-driven controllers treated as black-box functions without further interpretable content. 
Thus, observable-based consistency alone is insufficient.

We propose to extend observable-based consistency with behavioral consistency relations for black-box components. These relations are defined in terms of observable metrics of the input-output behavior of the system. As the shielded policy \(\polshield\) is learned with the safety shield in the loop, its behavior naturally reflects the interaction between the agent and the shield, and we can capture their consistency through behavioral metrics.
\begin{definition}[$\cragent$]\label{def:cragent}
    Let \(J_1, \ldots, J_K\) be \(K\) behavioral metrics and \(\thr_i\) their thresholds.
    The consistency relation $\cragent$ holds if all behavioral constraints for the shielded policy $\polshield$ are satisfied:
    \begin{equation}
    \cragent(c_\phi, \pi_\theta) \Longleftrightarrow
    \forall i = 1,\dots,K:\; J_i(\polshield) \geq \thr_i.
    \end{equation}
\end{definition}
Each metric $J_i$ captures a relevant aspect of system behavior, and the corresponding threshold $\thr_i$ specifies the minimal acceptable performance. This formulation allows consistency to be assessed independently of the internal representation of the policy, and enables detailed analysis of violations. Moreover, the task-specific behavioral metrics can be easily extended, and it is straightforward to trace consistency violations back to a specific metric.

\subsection{Consistency Preservation Rules}\label{sec:cprs}
The consistency relations $\crshield$ and $\cragent$ identify when a \ac{V-SUM} is inconsistent, but do not specify how consistency can be restored. 
A key aspect of our setting is the asymmetry between the structured models and the data-driven components. While the safety shield can be modified by standard model transformations to update its parameters, the data-driven controller is changed indirectly through retraining. We describe these two cases separately.
Changes in other models of the V-SUM, e.g., the physical parameters, induce changes to the parameter \(\phi\) of the safety shield.
\begin{definition}[$\cprshield$]
    The consistency preservation rule \hfill\break
    $\cprshield$ maps changes in the source models to updated parameters $\phi' = \delta_{\Mshield}(\phi)$ of the safety shield such that
	\begin{equation}
		\phi' \in \bigcap_{i=1}^k o_i ( m_i' ),
		% \phi' \subseteq \bigcap_{i=1}^k o_i \big( \delta_{M_i}(m_i) \big),
	\end{equation}
	where $m_i' = \delta_{M_i}(m_i)$ are the updated model instances.
    % \commentRomain{What about writing is less formally? Something like "The consistency preservation rule $\cprshield$ maps changes in the source models to an updated parameter set $\phi'$ of the safety shield such that [equation], where $m_i' = \delta_{M_i}(m_i)$ are the updated model instances.}
\end{definition}
Intuitively, $\cprshield$ changes the parameters of the safety shield such that they remain consistent with the updated constraints imposed by the other models' changes. This is essentially a standard model transformation, where the updated shield parameters are obtained directly from the observables of the modified models. Note that the update is unidirectional here: we only encode how changes in other models affect the safety shield but do not allow for changes in the safety shield to induce changes in the other models.

In contrast, changes in the data-driven controller cannot be expressed as a direct transformation of its parameters. Since the policy \(\safepol\) is a black box, consistency can only be restored through a learning process (i.e., retraining the agent) that enforces the behavioral constraints defined by \(\cragent\) (Def.~\ref{def:cragent}).
\begin{definition}[$\cpragent$]
    The consistency preservation rule \hfill\break
    $\cpragent$ returns updated parameters $\theta' = \delta_{\Magent}(\theta)$ such that
    \begin{equation}
    \forall i = 1,\dots,K:\; J_i(\tilde{\pi}_{\theta'}) \geq \thr_i.
    \end{equation}
\end{definition}
Based on the extent of the changes, various strategies can be used (with increasing computational costs):
\begin{enumerate}
    \item Fine-tuning the existing agent (its policy) with the updated safety shield on small amounts of data,
    \item Retraining the agent from scratch with the updated safety shield, or
    \item Redesigning a new architecture and training it from scratch with the updated safety shield and updated hyperparameters.
\end{enumerate}

The two \ac{CPR}s further emphasize the asymmetry between \(\crshield\) and \(\cragent\) since $\cprshield$ is essentially a deterministic model transformation but $\cpragent$ relies on a learning process based on behavioral feedback. Besides that, restoring consistency for data-driven components is inherently approximate and depends on the specified performance thresholds in Def.~\ref{def:cragent}. With this distinction, we propose a general approach for integrating black-box components into model-driven engineering: since their internal structure is not accessible, consistency is defined at the behavioral level and assessed and restored using metrics over system executions.

%% file: figures/overview.tex
\begin{tikzpicture}[scale=1.0]
	\tikzstyle{cr}  = [draw, line width=1.0, <->, >=latex]
	\tikzstyle{io}  = [draw, dashed, line width=1.0, ->, -stealth]
	\tikzstyle{cpr} = [draw, dashed, line width=1.0, ->, >=latex]
	\def\umltabw{0.4cm}   % tab width
	\def\umltabh{0.2cm}  % tab height
	\tikzset{
		umlpkg/.style={
				rectangle, draw=#1, line width=1.0,
				fill=#1, fill opacity=0.3, draw opacity=1, text opacity=1,
				align=center, inner sep=6pt, outer sep=0pt, minimum height=0.8cm,
				append after command={
						\pgfextra{%
							\coordinate (umlbl) at (\tikzlastnode.north west);
							\fill[#1, fill opacity=0.3] (umlbl) rectangle ++(\umltabw,\umltabh);
							\draw[line width=1.0, draw=#1, line cap=butt]
							(umlbl) -- ++(0,\umltabh) -- ++(\umltabw,0) -- ++(0,-\umltabh);
						}%
					}
			}
	}
    \tikzstyle{every node}=[font=\small]
    \tikzstyle{othermodel} = [umlpkg=gray, minimum width=1.1cm, minimum height=0.6cm]

    \draw[loosely dashed, rounded corners] (-3.5, -2.8) rectangle (4.8, 0.9);
    \node[below left] at (4.7, 0.9) {\ac{V-SUM} \(M\)};

	% \node[umlpkg=darkblue, minimum width=\minwidth] (shield) {Safety shield $c_\phi$};
	\node[umlpkg=darkblue] (shield) {Safety shield $c_\phi$};
	\node[umlpkg=darkblue, left=0.9cm of shield.west] (shield_meta) {\(\Mshield\)};
	% \node[umlpkg=darkgreen, minimum width=\minwidth, below=1.4cm of shield.south] (agent) {Agent $\pi_\theta$};
	\node[umlpkg=darkgreen, below=1.4cm of shield.south] (agent) {Agent $\pi_\theta$};
	\node[umlpkg=darkgreen, left=1.3cm of agent.west] (agent_meta) {\(\Magent\)};

	\node[right=1.6cm of shield.east] (collection) {};
	\node[othermodel, below right=0.75cm and 0.6cm of shield.east] (m1) {$m_1$};
	\node[othermodel, below=0.6cm of m1.south] (M1) {\(M_1\)};
	\node[othermodel, right=0.4cm of m1.east] (m2) {$m_2$};
	\node[othermodel, below=0.6cm of m2.south] (M2) {\(M_2\)};
	% \node[umlpkg=gray, minimum width=2.9cm, below right=1.8cm and 0.6cm of collection.east]
	% (m3) {Other model \\ $m_3 \in M_3$};

	% Lines
    \draw[io] (shield.west) -- node[midway, above] {<<i.O.>>} (shield_meta.east);
    \draw[io] (agent.west) -- node[midway, above] {<<i.O.>>} (agent_meta.east);
    \draw[io] (m1.south) -- node[midway, right] {<<i.O.>>} (M1.north);
    \draw[io] (m2.south) -- node[midway, right] {<<i.O.>>} (M2.north);

	\draw[cr] (shield.south) -- node[midway, above left=-0.1cm and 0.0cm] {$\cragent$} (agent.north);
	\draw[cr] (shield.east) -| node[midway, below right=-0.0cm and 0.1cm] {$\crshield$} (m1.north);
	\draw[cr] (shield.east) -| (m2.north);

    \draw[cpr, purple] ($(shield.south) + (0.2,0.0)$) -- node[right] {$\cpragent$} ($(agent.north) + (0.2,0.0)$);
	\node[purple, below right=0.0cm and 0.2cm of shield.south, align=center, font=\footnotesize] {$\delta(\phi)$};
	\node[purple, above right=0.0cm and 0.2cm of agent.north, align=center, font=\footnotesize] {$\delta_{\Magent}(\theta)$};

    \draw[cpr, orange] ($(m2.north) + (0.2,0.0)$) |- node[midway, above left = 0.0cm and 0.4cm] {$\cprshield$} ($(shield.east) + (0.0,0.2)$);
	\node[orange, above right=0.0cm and 0.15cm of m2.north, align=center, font=\footnotesize] {$\delta(m_2)$};
	\node[orange, above right=0.2cm and 0.1cm of shield.east, align=center, font=\footnotesize] {$\delta_{M_\text{shield}}(\phi)$};
\end{tikzpicture}

%% file: sections/experiments.tex
We demonstrate the applicability of our framework on a reach-avoid task for a simple \ac{CPS}.
We investigate how changes to the underlying system affect the shielded data-driven controller, while the \ac{V-SUM} is in different states of consistency. 

\subsection{System and Safety Shield}\label{sec:system}
We consider a unicycle system with discrete-time dynamics and no system noise, simulated by semi-implicit Euler integration:
\begin{equation}\label{eq:unicycle_dynamics}
	x_{k+1} = 
    \begin{bmatrix}
        p_{k+1} \\
        v_{k+1} \\
        \alpha_{k+1}
    \end{bmatrix} = 
    f(x_k, u_k) = 
	\begin{bmatrix}
		p_k + \Delta t\, v_{k+1}\, e_{\alpha_{k+1}} \\
		v_k + \tfrac{\Delta t}{m} F_k \\
		\alpha_k + \Delta t\, \omega_k
	\end{bmatrix},
\end{equation}
where \(e_\alpha = [\cos \alpha, \sin \alpha]^\top\) and \(m\) is the mass of the system.
The system state is $x = [p_x, p_y, v, \alpha]^\top$, where $p_x$ and $p_y$ are the agent position, $v$ is the longitudinal velocity, and $\alpha$ is the yaw angle.
The control input is $u = [F, \omega]^\top$, where $F$ is the longitudinal force and $\omega$ is the yaw rate. The input space $\mc{U} = [-F_b, F_a] \times [-\omega_\text{max}, \omega_\text{max}]$ is defined through the maximum braking force and accelerating force, $F_b$ and $F_a$, and the max yaw rate $\omega_\text{max} = \pi / 2$. The environment consists of 10 circular obstacles with positions \([o_{i,x}, o_{i,y}]^\top\) and radii \(r_i\), and a goal state \([g_x, g_y]^\top\).

Safety is specified with one discrete control barrier function per obstacle, which ensures that the distance to the obstacle \(i\) is larger than the braking distance towards it:
\begin{equation}\label{eq:dcbf}
	h_i(x) = \underbrace{\Vert p-o_i \Vert_2 - r_i}_{\text{distance to boundary}} - \underbrace{\frac{m \, v^2_{\text{app},i}}{2F_b}}_{\text{braking distance}} \geq 0,
\end{equation}
where \(v_\text{app,i} = v (n_i^\top e_\alpha)\) is the velocity towards the obstacle center, and \(n_i
\) is the normal vector pointing from the agent towards the center of the obstacle. The set of safe states is 
\(\mc{X}^s =\{x \in \mc{X} \mid h_i(x) \geq 0,\; \forall i=1,\dots,N\}\).
The inputs \(u_k = \pi_\theta(x_k)\) of the \ac{RL} policy \(\pi_\theta \in \Magent\) are corrected by the safety shield via boundary projection (see Sec.~\ref{sec:pre:acrl}), such that the next state satisfies \(h_i(x) \geq 0\). This amounts to solving the optimization problem~\cite[Sec.~II-C]{agrawal_discrete_2017}: 
\begin{equation}\label{eq:safety_shield}
	\begin{aligned}
		\tilde{\pi_\theta}(x_k) = & \arg \min_{\tilde{u}} \Vert u_k -\tilde{u} \Vert_2 \\
		\text{s.t.} \quad         & h_i(f(x_k, \tilde{u})) \geq (1 - \gamma) h_i(x_k) \, \forall i, \text{ and } \tilde{u} \in \mc{U}, %\\
		                          % & \tilde{u} \in \mc{U},
	\end{aligned}
\end{equation}
where \(\gamma \in (0, 1]\) defines the rate of convergence to the safe set. With this safety shield formulation, the safe input set \(\safeu(x_k)\) is implicitly defined by the constraints in the optimization problem. Since the unicycle dynamics in \eqref{eq:unicycle_dynamics} are non-linear, we linearize \(h_i(f(x_k, \tilde{u}))\) in \(u\) with a first-order Taylor approximation to obtain a convex quadratic problem,
and tighten each constraint by a margin that accounts for the linearization error~\cite{agrawal_discrete_2017}.

\subsection{Reinforcement Learning Controller}

The controller $\pi_\theta \in \Magent$ is represented by a fully connected neural network with two hidden layers and ReLU activations and trained with \ac{PPO} \cite{schulman_proximal_2017}.
The reward encourages progress toward the goal, measured by the distance \(d_k\), while penalizing collisions and excessive acceleration:
\begin{equation}
	\begin{aligned}
		 r(&u_k, x_k) =                                                                                                                  \\
		 & \begin{cases}
			   10,                                                                         & \text{if the goal is reached}, \\
			   -10,                                                                       & \text{if a collision occurred}, \\
			   \frac{d_{k-1} - d_k}{3 \dt} - \dt \, \max\{0, F \} - 0.5 & \text{otherwise}.
		   \end{cases}
	\end{aligned}
\end{equation}
An episode lasts until the goal is reached, a collision occurs, or a maximum episode length of 150 steps is reached. At the start of each episode, the agent position, goal position, and the position and radii of the obstacles are initialized randomly.
Hyperparameters are selected through Bayesian optimization over 100 runs, after which the
five best configurations (see the Appendix) are each trained for two million environment steps, yielding five
shielded policies $\polshield$.

\subsection{Consistency Relations}
As the problem is a reach-avoid task and the controller is trained with \ac{RL}, both the goal reaching rate and the expected cumulative discounted reward \(J(\polshield)\) are natural choices of metrics for \(\cragent\). Our safety shield in \eqref{eq:safety_shield} only overrides, when the proposed input \(\pi_\theta(x)\) is outside \(\safeus\). This allows using the intervention rate:
\begin{equation}\label{eq:intervention_rate}
	\text{IR}(\tilde{\pi}_\theta) = \mathbb{E}_{x \sim d^{\polshield}} \left[ \mathbb{I} \big( \pi_\theta(x) \notin c_\phi(x) \big) \right],
\end{equation}
where $d^{\polshield}$ is the state distribution under the shielded policy $\polshield$, and $\mathbb{I}$ is the indicator function. A high intervention rate indicates a misalignment between the safety shield and the agent.
Instead of defining thresholds $\tau_i$, we apply $\cpragent$ (Sec.~\ref{sec:experiments:cpr}) after every system change and assess which metrics are suitable for $\cragent$.

The only other model of the \ac{V-SUM} is the \ac{RL} environment $m_\text{env}$, which implements the dynamics in \eqref{eq:unicycle_dynamics}. 
The safety shield for the unicycle system is parameterized as
\begin{equation}
	\phi = \left([-F_b, F_a] \times [-\omega_\text{max}, \omega_\text{max}], f, \mc{X}^s \right),
\end{equation}
where \(f\) depends on \(m\) and \(\mc{X}^s\) depends on \(F_b\) and \(m\) through \eqref{eq:dcbf}.
The observable $o_\text{env}(m_\text{env}) = \{\phi_\text{env}\}$ returns the shield parameters implied by the environment, so $\crshield$ holds if and only if $\phi = \phi_\text{env}$, i.e., the shield and the environment agree on $F_b$, $F_a$, and $m$.

\subsection{Consistency Preservation Rules}\label{sec:experiments:cpr}
The change $\delta_{M_\text{shield}}(\phi)$ induced by $\cprshield$ updates the input bounds $\mathcal{U}$ with $F_a$ and $F_b$, and the constraints in \eqref{eq:safety_shield} with $m$ and $F_b$, as given by the environment observable \(o_\text{env}(m_\text{env})\).
For $\cpragent$, we consider two repair strategies.
First, $\delta_{\Magent}^\text{fine} (\theta)$ fine-tunes the trained policy $\tilde{\pi}_{\theta}$ for an  additional $2 \times10^5$ environment steps with a halved learning rate. Second, $\delta_{\Magent}^\text{re} (\theta)$ retrains the policy from scratch on $2 \times10^6$ environment steps using the same hyperparameters.

\subsection{Experimental Setup}
\begin{figure*}[tb]
	\centering
	\includegraphics[scale=0.7]{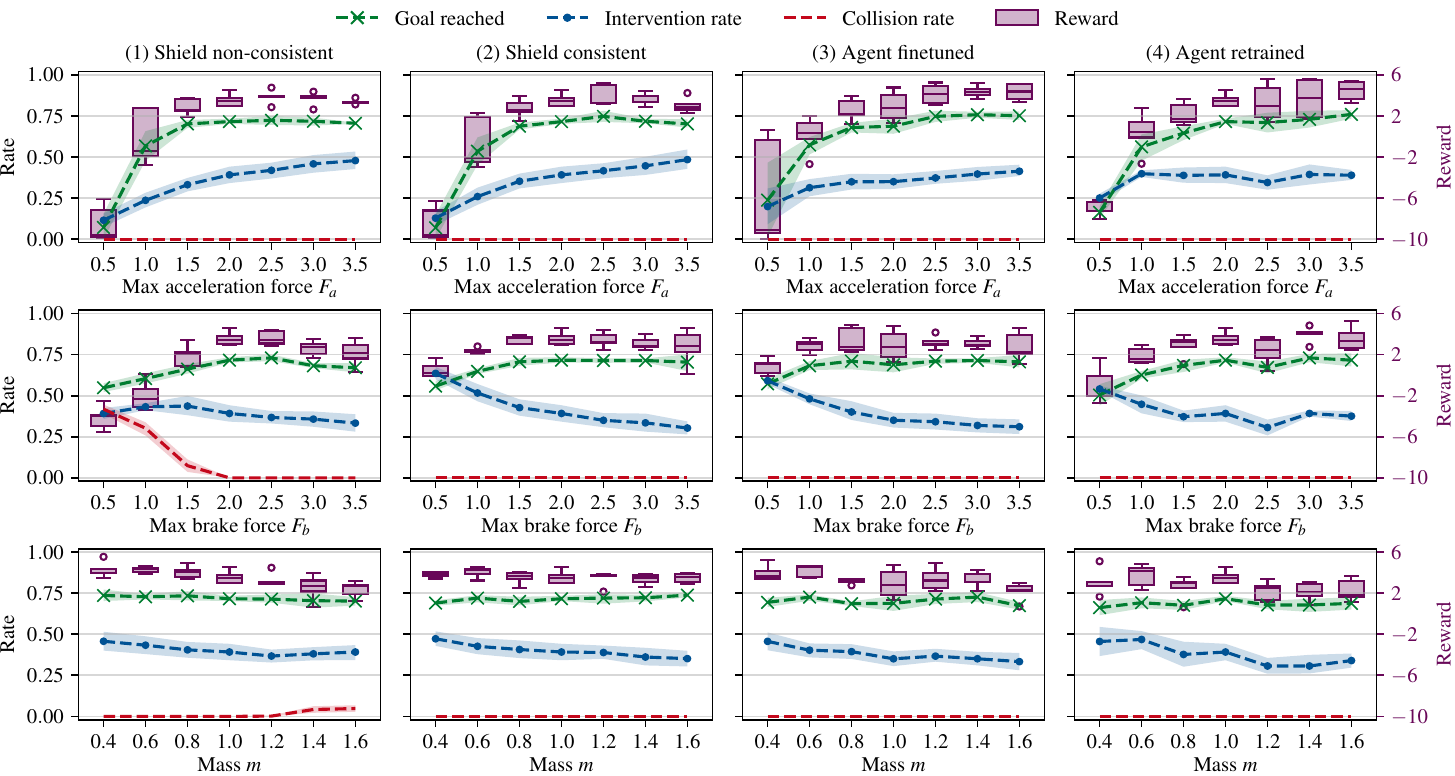}
	\caption{The performance of the shielded policy $\polshield$ on 100 episodes per configuration across the parameter sweep and different states of consistency. The boxes summarize the five best hyperparameter configurations, each averaged over 100 episodes, and the shaded bands are the 95\% confidence intervals.}
	\label{fig:results}
	\Description{Results.}
\end{figure*}

We train the shielded agent with \emph{initially consistent values $F_b=F_a=2$, $m=1$}. The parameters $F_b$ and $F_a$ are varied in increments of $0.5$ from $0.5$ to $3.5$, and the mass $m$ in increments of $0.2$ from $0.4$ to $1.6$. For each parameter setting, we evaluate the shielded policy on four different scenarios: 
\begin{enumerate}
	\item\label{item:state:1} No \ac{CPR} applied, i.e., $\crshield$ is violated,
	\item\label{item:state:2} $\crshield$ is satisfied after applying $\delta(\phi)$,
	\item\label{item:state:3} Applying $\delta_{M_\text{shield}}(\phi)$ and $\delta_{\Magent}^\text{fine} (\theta)$,
	\item\label{item:state:4} Applying $\delta_{M_\text{shield}}(\phi)$ and $\delta_{\Magent}^\text{re} (\theta)$.
\end{enumerate}
In each configuration, we deploy each of the five shielded policies (trained with the five best hyperparameter configurations of the Appendix) for 100 episodes and report the reward, goal reaching rate, collision rate, and intervention rate from~\eqref{eq:intervention_rate} in Fig.~\ref{fig:results}. 
Our code is available at \url{https://anonymous.4open.science/r/consistency-shielded-rl-2047/}. 

\subsection{Results and Discussion}

\paragraph{Shield inconsistency (scenario~\ref{item:state:1})}
Without updating the shield, collisions occur primarily for low braking forces and high mass values, confirming that outdated shield assumptions invalidate the safety guarantees.
These collisions \emph{emphasize the importance of ensuring the consistency of the safety shield} with system changes.

\paragraph{Updating the shield (scenario~\ref{item:state:2})}
Repairing the shield parameters eliminates collisions across the parameter sweep, confirming that satisfying $\crshield$ restores safety. However, the repair also affects task performance.
For low $F_b$, the longer braking distance in \eqref{eq:dcbf} makes the shield more conservative, so it intervenes more often.
For high $F_b$, the shorter braking distance gives the agent more freedom and improves the goal reaching rate.

\paragraph{Fine-tuning (scenario~\ref{item:state:3}) and retraining (scenario~\ref{item:state:4}) the controller}
Both fine-tuning with $\delta_{\Magent}^\text{fine} (\theta)$ and retraining with $\delta_{\Magent}^\text{re} (\theta)$ improve the reward and goal reaching rate at higher \(F_a\) compared to scenario~\ref{item:state:2}, where higher \(F_a\) slightly decreases performance. However, in other configurations the effect of the two \ac{CPR}s is minor.

\paragraph{Behavioral consistency metrics}
The results also inform the choice of metrics for $\cragent$.
Lower \(m\) and higher \(F_a\) exhibit a higher intervention rate, but similar, or improved reward and goal reaching rate, suggesting that increased interventions can be compatible with improved task performance, which is consistent with recent work~\cite{markgraf_safe_2026}. A threshold on the median episode reward is more informative and identifies configurations with degraded performance in most cases. However, in \ac{RL}, the agent can exploit the reward in unintended ways, hence a high reward does not always mean good performance. 
Behavioral consistency of a learned controller is thus best captured by a combination of task specific metrics.

%% file: sections/conclusion.tex
We present a framework for consistency-based change evaluation of shielded data-driven controllers in \ac{CPS}. By treating the safety shield and the data-driven controller as first-class models within a \ac{V-SUM}, we define explicit safety and behavioral consistency relations that keep guarantees valid under model changes. The consistency of the safety shield is assessed through observables on other models. Since data-driven controllers are opaque, behavioral consistency is assessed over system-level metrics rather than internal structure. Although our evaluation targets an \ac{RL} controller, the framework applies to any black-box component whose consistency can only be assessed at the behavioral level.

Most notably, our evaluation on a reach-avoid task highlights the importance of including the safety shield in multi-model consistency management for autonomous \ac{CPS}.
Changes to the system during design or through wear can invalidate the assumptions of the shield and lead to collisions. Checking and repairing the consistency of the shield prevents this.
The results further indicate that general metrics (e.g., reward) should be combined with task-specific metrics (e.g., goal reaching rate) to assess behavioral consistency.
The performance degradation after a shield repair, and thus the benefit of fine-tuning or retraining as \ac{CPR}s, was minor in many configurations.
We partly attribute this to the simplicity of the unicycle system and expect larger effects for more complex systems and for controllers that operate closer to their performance limits.
Future work should investigate such settings, and whether more sophisticated fine-tuning techniques, such as layer freezing or tailored learning-rate scheduling, recover performance reliably.

%% file: sections/appendix.tex
The hyperparameters used in our evaluation are listed in Table~\ref{tab:hyperparameters}.

\begin{table}[H]
	\centering
	\caption{PPO hyperparameters of the five best configurations from the Bayesian optimization with 100 runs.}
	\label{tab:hyperparameters}
	\begin{tabular}{lccccc}
		\toprule
		Hyperparameter & C1 & C2 & C3 & C4 & C5 \\
		\midrule
		Learning rate ($\times 10^{-4}$) & 7.06 & 5.72 & 7.27 & 5.32 & 4.45 \\
		Batch size & 256 & 256 & 128 & 128 & 256 \\
		% Epochs per update & 3 & 3 & 5 & 4 & 4 \\
		Clip range $\epsilon$ & 0.104 & 0.145 & 0.101 & 0.104 & 0.123 \\
		Entropy coefficient & 0.029 & 0.021 & 0.013 & 0.021 & 0.020 \\
		Vf. coefficient & 0.513 & 0.520 & 0.407 & 0.543 & 0.349 \\
		Hidden layer size & 512 & 512 & 512 & 512 & 512 \\
		\bottomrule
	\end{tabular}
\end{table}